\documentclass[aps,pra,showpacs, oneside,preprintnumbers, amsmath,twocolumn, superscriptaddress, 9pt]{revtex4-1}

\usepackage{amsmath}
\usepackage{amssymb}
\usepackage{graphicx}
\usepackage{multirow}
\usepackage{epstopdf}
\usepackage{booktabs}

\usepackage{dcolumn}
\usepackage{bm}
\newcommand{\ignore}[1]{}

\begin{document}

\def\e{\mathcal{E}}

\title{EIT Intensity Correlation Power Broadening in a Buffer Gas}

\author{Aojie Zheng}
\affiliation{Lewis \& Clark College, Portland, OR 97219}

\author{Alaina Green}
\affiliation{Lewis \& Clark College, Portland, OR 97219}

\author{Michael Crescimanno}
\affiliation{Department of Physics and Astronomy, Youngstown State
University, Youngstown, OH 44555-2001}

\author{Shannon O'Leary} 
\affiliation{Lewis \& Clark College, Portland, OR 97219}

\date{\today}

\begin{abstract}

EIT noise correlation spectroscopy holds promise as a simple, robust 
method for performing high resolution spectroscopy used in  
optical magnetometry and clocks. Of relevance to these applications, 
we report here on the role of buffer gas pressure and magnetic field gradients  on power broadening of Zeeman EIT noise correlation resonances.

\end{abstract}





\maketitle

\section{INTRODUCTION}
\label{sec:intro}

When laser fields interact with atomic vapors to induce atomic coherence between dipole-forbidden quantum states, such as between two ground states in a three-level $\Lambda$ system, the optical properties of the atomic vapor are dramatically altered in controllable and useful ways.  Simultaneously, interactions with the atomic vapor modifies the participating light fields.  A well-known representative coherent phenomenon of this type is electromagnetically induced transparency (EIT) \cite{EITreview}, which owes its spectrally narrow transmission window and ultra-steep dispersion to optical pumping into a dark state, which results from quantum interference between transition pathways.  The controllable optical properties that accompany EIT and related coherent phenomenon such as coherent population trapping (CPT) and electrically induced absorption (EIA) are particularly attractive for applications \cite{Lukinreview} such as atomic clocks \cite{CPTreview}, magnetometry \cite{budker}, quantum information and communication schemes \cite{vanderWalScience, duan2001nature} and quantum computation \cite{knill_nature_computation}.  This widespread interest is motivation for understanding properties of the light fields emerging from interaction with atomic coherence.

Although phase noise is not easily detectable by photodetectors, light-matter interactions like typical atomic absorption or EIT map the phase noise onto the light's amplitude (and thus, intensity) fluctuations.   These phase modulation to amplitude modulation (PM $\rightarrow$ AM) conversion processes \cite{Camparo, Zoller} often produce amplitude noise greater than the laser source's residual intensity noise (RIN). 

PM $\rightarrow$ AM conversion processes impart intensity noise with spectroscopic information, making it a useful spectroscopic tool \cite{PRL1991, McIntyre, rosenbluh}.  Monitoring correlations between the coherence-derived light fluctuations in the two EIT (or EIA) laser fields using an intensity cross-correlation noise statistic referred to as $g^2(0)$, a.k.a.\ EIT noise correlation spectroscopy, shows great promise for a range of applications due to a narrow, power-broadening resistant resonance described below and studied previously in \cite{Scully, YanhongPRL, Felinto:13, Lee:13}. 

Intensity fluctuations of the two EIT laser fields become strongly correlated on EIT resonance, and abruptly strongly anticorrelated slightly off-resonance.  Correlation-anticorrelation switching near the EIT resonance gives rise to a narrow noise correlation resonance, whose linewidth is up to an order of magnitude more narrow than the underlying EIT transparency peak, and has been shown in some cases to be more narrow that the zero-power EIT transit-limited lifetime \cite{YanhongPRL}.  
  
This narrow correlation resonance is of further scientific and technological interest because it has been shown in experiment and in theory models to be power-broadening resistant at laser intensities which broaden the underlying EIT resonance \cite{Scully, harvard2}.  Coherence-based applications that exploit this correlation resonance will have an advantage in sensitivity over those with the usual optical schemes, and for many applications, the prospect of working at higher optical powers without the sensitivity-defeating consequences of broadening is clearly technologically attractive.   More recent experiments indicate that the noise correlation resonance does eventually broaden at higher laser intensities, but questions remain about the physical causes of the broadening.

A recent paper \cite{Xiao_2014} presents an experimental and theoretical study of factors that affect the linewidth of the EIT intensity correlation resonance in a vacuum cell.  The experiments and analysis presented there reveal linewidth dependencies on laser power when full anticorrelation of the intensity fluctuations has not been reached, most prominently linewidth narrowing with increasing laser power, down to a minimum of $25$ kHz when a narrow linewidth (1 MHz) laser is used, and linewidth broadening with increasing power when a non-narrow linewidth laser (80 MHz) is used.  The authors attribute the linewidth broadening to the influence of the laser's RIN.

In this article, we report on results from new experiments on intensity fluctuations of orthogonally polarized fields in the EIT Hanle configuration of the ${}^{87}$Rb  $D_1$ line. This work extends our understanding of EIT intensity correlations to buffer gas cells.  It is well-known that introducing a buffer gas prolongs $\Lambda$-type dark state coherence lifetimes by increasing the atomic transit time via collisional diffusion.  In systems where transit time broadening dominates over collisional broadening, dark state resonance linewidths such as CPT or EIT become more narrow as buffer gas pressure is increased \cite{PhysRevA.56.R1063}.  We have chosen these buffer gas cells to exploit the longer EIT lifetimes and help identify physical processes that affect the EIT noise correlation power broadening slope.

To distinguish from earlier vacuum cell experiments, our work with buffer gas cells show EIT-derived correlation linewidths of less of than 1 kHz are easily obtained with a non-narrow linewidth (55 MHz) laser, and full correlation of the intensity fluctuations can be reached at higher laser powers.  
  
We also show that buffer gas pressure and magnetic field gradients play significant roles in determining the power broadening slope of Zeeman-EIT noise correlation resonances.
In particular, in Section \ref{sec:results} we report experimental results and numerical simulations that simultaneously show the expected relationship between EIT linewidth and buffer gas pressure, and a counter-intuitive, reverse relationship between the linewidth of the EIT-derived intensity correlation resonance and buffer gas pressure.  These new experimental and numerical results are accompanied by a qualitative discussion of underlying physics.  In Section \ref{sec:gradients} we report on a new experiment and accompanying theory model that describes the role of magnetic field gradients on the power broadening slope of the narrow correlation resonance.  Finally, we summarize and contextualize our work in the concluding section.
 
\section{Experimental Setup}
\label{sec:exp}

\begin{figure}
\includegraphics[width=\linewidth]{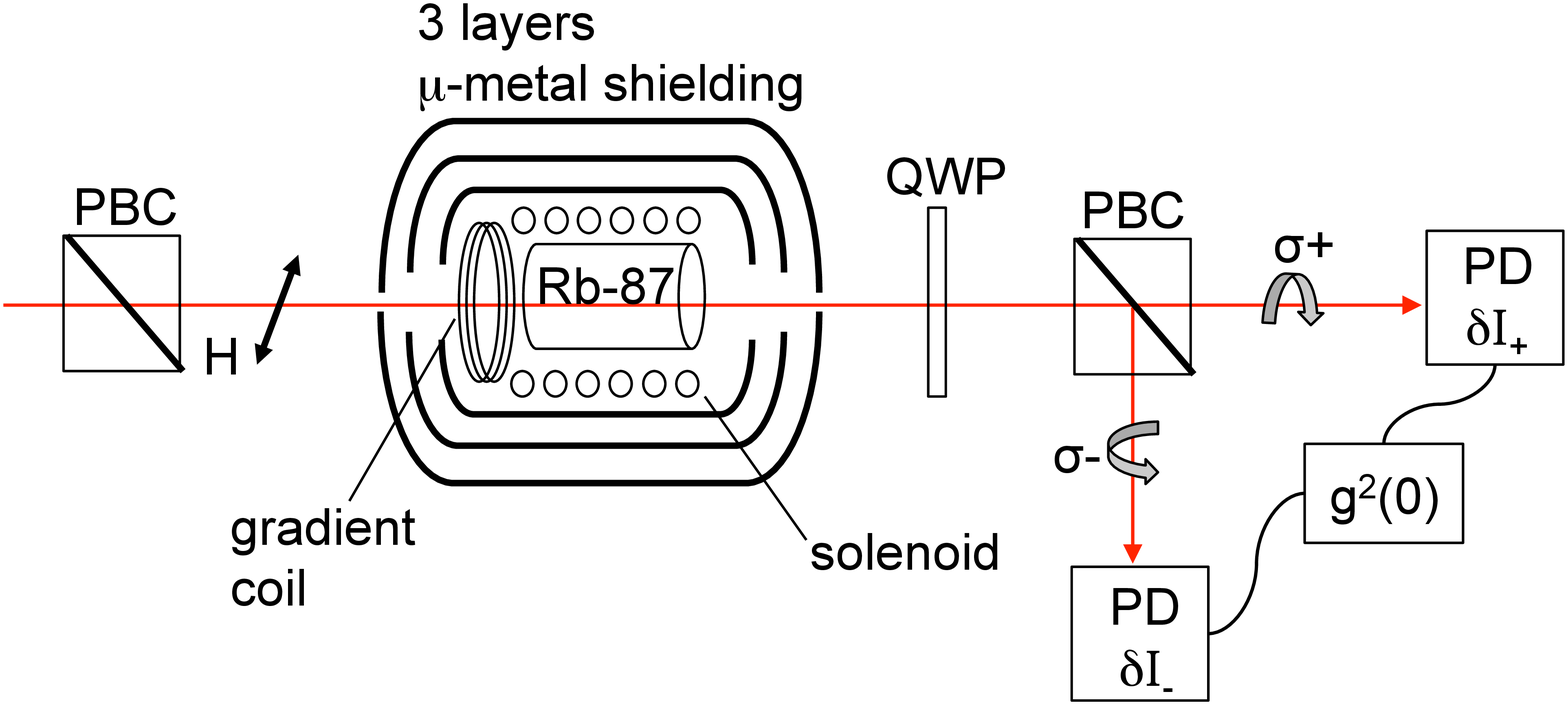}
\caption{(color online) Experimental setup.} 
\label{setup}
\end{figure}

A schematic of our experimental setup is shown in Fig.\ \ref{setup}.  We induce Zeeman EIT in a warm ${}^{87}$Rb buffer gas vapor cell using the Hanle configuration, which employs orthogonal circularly polarized components $\sigma+$ and $\sigma-$ of a single linearly polarized laser to induce an atomic coherence between degenerate ground state Zeeman sublevels.  The vapor cell is heated to  $\sim 50^\circ$C, and contains a buffer gas of either 2 Torr Ne or 10 Torr (5 Torr Ne + 5 Torr Ar) with a cell length of $l$ = 5 cm or 8 cm, respectively.  In order to observe Zeeman EIT, the atomic vapor cell is  shielded from the ambient magnetic field of the laboratory with three nested layers of $\mu$-metal shielding.  A nearly homogeneous magnetic field is created by a solenoid inside the innermost shield layer.  This longitudinal B-field is co-linear with the beam's propagation direction and is used to split the ground state Zeeman sublevels, and thus to control the two-photon detuning, $\Delta$. A small coil mounted far from the cell is used in the experiments described in Section \ref{sec:gradients} to introduce an axial magnetic field gradient across the vapor cell.

Using temperature and current control, we tune a commodity 795 nm non-narrow (55 MHz) linewidth free-running diode laser to the F=2 $\rightarrow$ F'=1 hyperfine transition of the ${}^{87}$Rb  $D_1$ line.  In the experiments shown in this paper, the laser beam diameter is 7 mm.  To ensure stability of the laser system against long-term temperature drifts in the lab environment, the center of the laser line can be locked to the transition using a dichroic atomic vapor laser lock (DAVLL) \cite{DAVLL}, though we have found that electronic locking is not strictly necessary. The large spectral bandwidth of such a ``noisy" laser is desirable for EIT noise correlation studies because its large intrinsic phase noise is much more relevant spectroscopically than the small remaining laser intensity instability \cite{zhang1995}. The RIN of our free-running laser diode is measured to be approximately $\sim0.2$\%.  This inexpensive diode laser was chosen for compatibility with device applications such as CPT atomic clocks \cite{kitching01a} and magnetometers \cite{CPT_magnetometer} and to demonstrate proof-of-principle that these simple lasers can provide a robust solution for in-the-field atomic-optical applications where external cavity alignment and stability are limited by the environment \cite{myneni}. 

The $\sigma+$ and $\sigma-$ components of the linearly polarized laser field couple two degenerate ground state Zeeman sub-levels to a common upper level, creating a three-level $\Lambda$ system. The resulting EIT is observed as a peak in the transmission of light as the atoms are subject to a magnetic field that is stepped through $B  \propto \Delta=0$. 

After the atomic vapor cell, a quarter waveplate and polarizing cube beamsplitter separate the $\sigma+$ and $\sigma-$ channels of the transmitted light, and each are individually detected with amplified photodiodes whose published bandwidths are limited to 1 MHz.  For each applied magnetic field, or two-photon detuning, $\Delta$, the intensity in each polarization channel is recorded at a rate of 1 MHz for 4 ms with enough dynamic range to detect the AC part, $\delta I_{\pm}$, of the photodiode signal, that is, the signal with the mean value subtracted. 

Of particular interest is the degree of correlation of the two fluctuating light fields, measured by the normalized intensity cross-correlation statistic, $g^2(0)$: 
\begin{equation}
   g^{2}(0)= 
\frac{\langle \mathrm{}(\delta
     I_{+})\,\mathrm{}(\delta
     I_{-})\rangle}{\sqrt{\langle(\mathrm{}\delta
       I_{+})^2\rangle\langle(\,\mathrm{}\delta I_{-})^2}\rangle}.
\label{eq:g2define}
\end{equation}

\noindent The numerator of $g^2(0)$ averages the products of the AC parts of the two signals, and the denominator normalizes the result such that perfect correlation outputs $g^2(0) = +1$ and perfect anticorrelation yields $g^2(0) = -1$.  The utility of this intensity cross-correlation statistic in probing EIT resonances was first reported in \cite{Scully}.

\begin{figure}
\includegraphics[width=\linewidth]{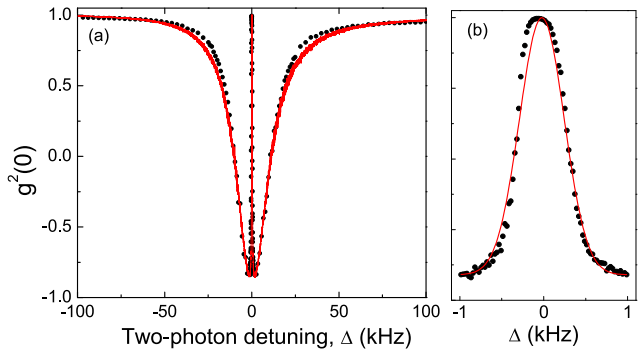}
\caption{(color online) (a) A representative $g^2(0)$ EIT intensity correlation curve.  Data points are the average of $g^2(0)$ values calculated from 10 consecutive data acquisition periods, and the (red) curve is from our theory model.  (b) The central correlation peak, with fitted Gaussian curve to determine FWHM.  These data were taken in a 10 Torr buffer gas cell with beam intensity $I=0.7$ mW/cm$^2$.}
\label{typical_data}
\end{figure}

Fig.\ \ref{typical_data}(a) shows a representative $g^2(0)$ dataset, which has a dual structure consisting of an outer basin that can approach complete anticorrelation, $g^2(0)=-1$, near zero-detuning and an inner central resonance peak arising from correlation switching to perfect correlation at zero two-photon detuning, $\Delta=0$.  The superimposed line in Fig.\ \ref{typical_data}(a) is from our theory model and is in good agreement with the experiment. We measure the central correlation peak's linewidth by fitting the experimental data to a Gaussian as shown in Fig.\ \ref{typical_data}(b).  

\section{Role of Buffer Gas pressure}
\label{sec:results}

\begin{figure}
\includegraphics[width=\linewidth]{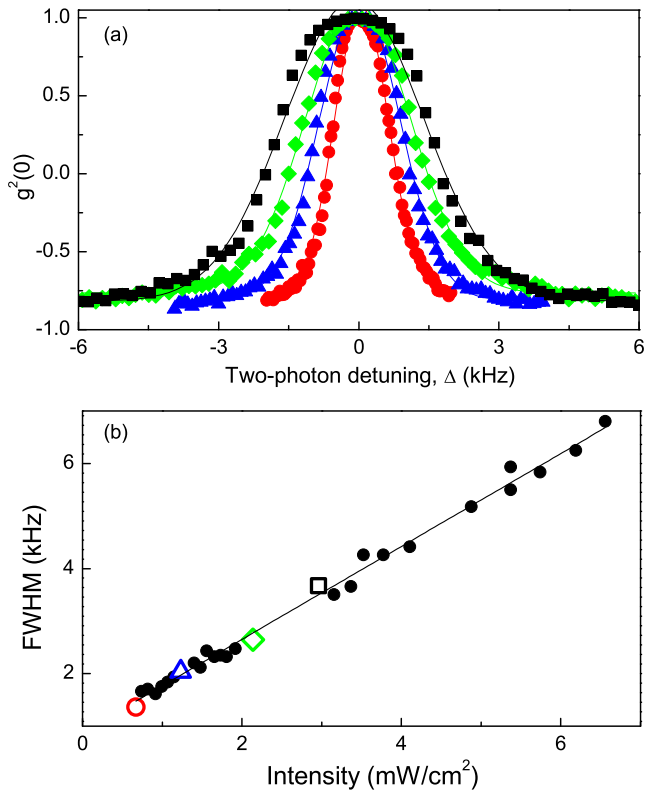}
\caption{(color online) (a) The $g^2(0)$ central correlation peak for four beam intensities.  (b) FWHM of the central correlation peak for a series of beam intensities.  Note that the FWHM of the peaks in (a) are indicated in (b) by corresponding open datamarkers.  These data were taken in a buffer gas pressure of 10 Torr.} 
\label{zerogradient_powerbroad}
\end{figure}

A series of experiments were performed to investigate the relationship of the central correlation peak's linewidth to laser power and buffer gas pressure.  Fig.\ \ref{zerogradient_powerbroad}(a) shows the central correlation peak broadening as laser power is increased in the 10 Torr cell described above.  Note that in each of the the traces shown in Fig.\ \ref{zerogradient_powerbroad}(a), the $g^2(0)$ value minimum approaches $-1$, and therefore the observed power-broadening cannot be attributed to curve-fitting artifacts that can arise when full anticorrelation is not reached and the central peak is not well-separated from the the outer basin.  

Fig.\ \ref{zerogradient_powerbroad}(b) summarizes the central peak's FWHM growth with laser intensity.  We observe that beyond a low power threshold, in this case $I>1$ mW/cm$^2$, the correlation linewidth broadens linearly with laser power.  We have defined the ``broadening rate" as the slope of this linear trend.  For the experimental conditions in Fig.\ \ref{zerogradient_powerbroad}, using buffer gas pressure of 10 Torr at 45 degrees, the broadening rate is 0.85 kHz per mW/cm$^2$.

The experimental result that the central EIT intensity correlation peak broadens with optical power is a challenge to the simple theories describing EIT-noise correlation, as their numerical evaluation does not include $\textit{a priori}$ predictions of this behavior.  In fact, the central noise correlation peak has been shown in experiment and theory to be impervious to broadening at low powers even as the underlying EIT peaks themselves suffer from power-broadening \cite{Felinto:13, harvard2}.  A relationship between the 
laser's RIN and power broadening of the correlation peak has been
established experimentally and modeled theoretically in \cite{YanhongPRL}.

As discussed in the introduction, buffer gas is used to lengthen interaction times and thus increase ground state coherence times, narrowing dark state resonances.  As one might expect, at low powers the underlying EIT peak narrows with buffer gas pressure and the EIT-derived noise correlation peak also narrows correspondingly.  Note that comparing $g^2(0)$ power-broadening in 2 Torr and 10 Torr buffer gas cells, we see that the ordering is reversed; at high laser powers, the noise correlation peak broadens with buffer gas pressure.

\begin{figure}
\includegraphics[width=\linewidth]{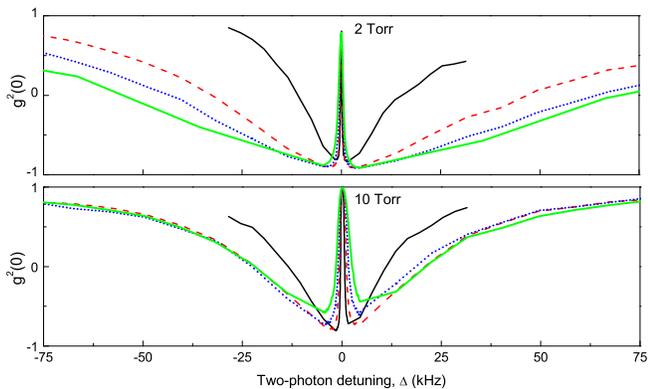}
\caption{(color online) Comparison between full $g^2(0)$ curves at 2 Torr (top) and 10 Torr (bottom) buffer gas pressures.  Thin solid (black): I $=0.6$ mW/cm$^2$; dashed (red): I $=1.2$ mW/cm$^2$; dotted (blue): I $=1.7$ mW/cm$^2$; and thick solid (green): I $=2.2$ mW/cm$^2$.} 
\label{g2_two_cells}
\end{figure}

In Fig. \ref{g2_two_cells} we show the $g^2(0)$ traces for buffer gas with 2 Torr and 10 Torr pressure, taken with otherwise identical experimental conditions.  As the laser intensity is increased, we observe the noted counter-intuitive result that the central correlation spike exhibits more broadening in the higher pressure cell than in the lower pressure cell.  We also show in Fig. \ref{g2_two_cells} that the outer correlation basin exhibits more power-broadening in the lower pressure cell.  This result is consistent with our expectations, since the outer basin scales with the underlying power-broadened EIT linewidth \cite{harvard2}.

\begin{figure}
\includegraphics[width=\linewidth]{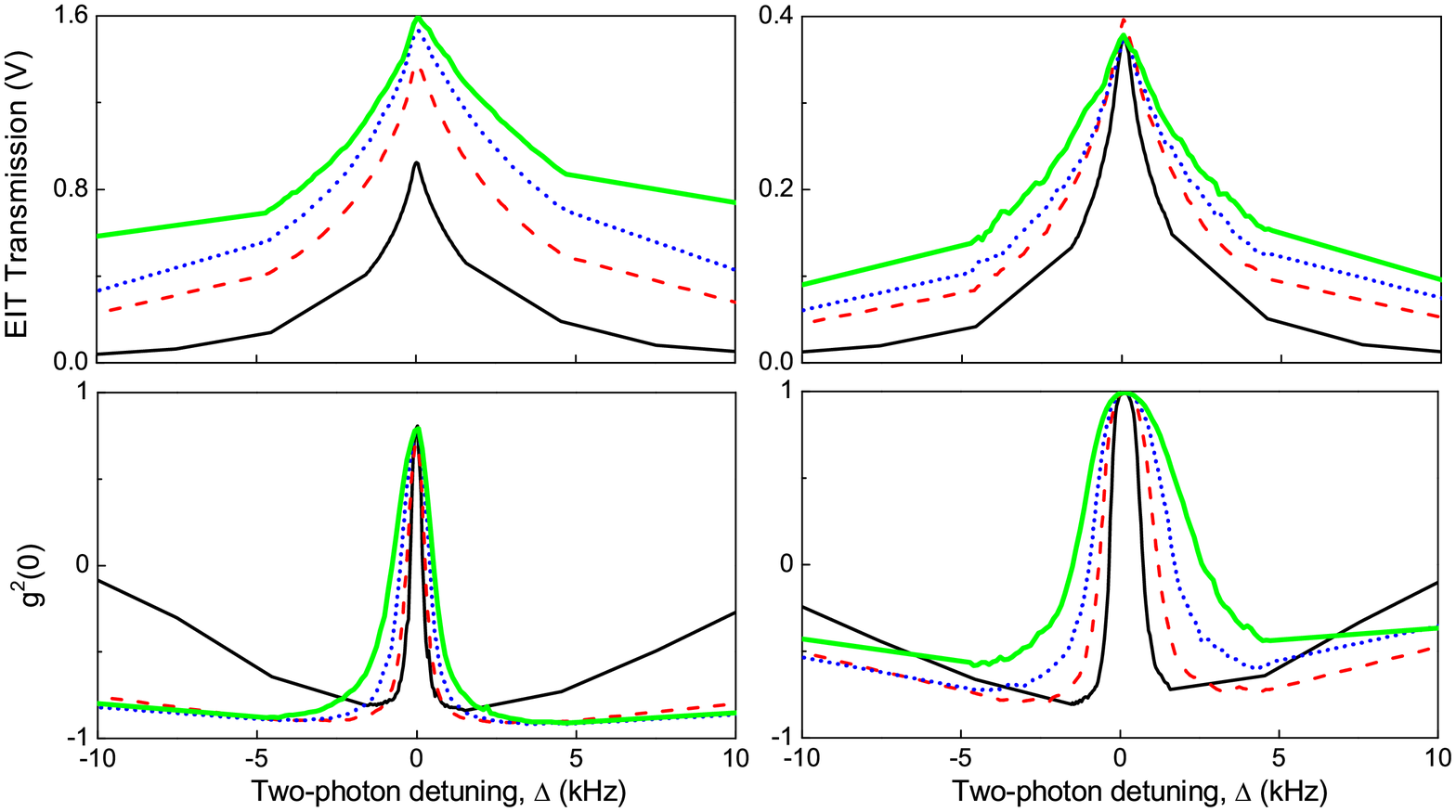}
\caption{(color online) Comparing EIT (top) and noise correlation resonances (bottom) in the 2 Torr (left) and 10 Torr (right) cells with optical powers in the power broadening regime.  Thin solid (black): I $=0.6$ mW/cm$^2$; dashed (red): I $=1.2$ mW/cm$^2$; dotted (blue): I $=1.7$ mW/cm$^2$; and thick solid (green): I $=2.2$ mW/cm$^2$.  EIT curve baselines have been subtracted so that they overlay at far detuning.  Note that in all cases, the width of each feature grows very nearly linearly with power, but that the noise correlation resonances broaden more readily in the cell that shows less EIT power broadening.} 
\label{EITcompare}
\end{figure}

\begin{figure}
\includegraphics[width=\linewidth]{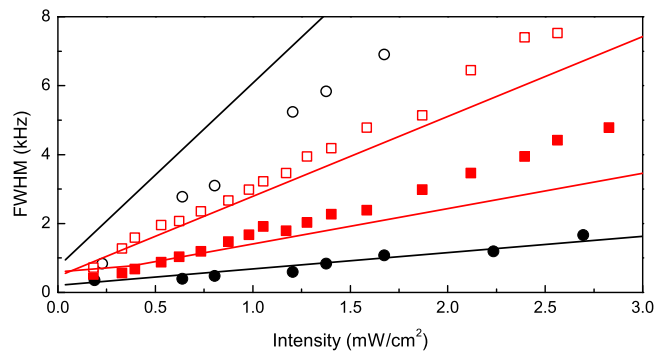}
\caption{(color online) Noise correlation peak FWHM (filled data markers) and EIT FWHM (open data markers) as a function of intensity for two buffer gas vapor cells, 2 Torr (circles, black) and 10 Torr (squares, red).  The lines are representative of the theory results, which reflect the same overall trends in the broadening slopes.} 
\label{summary graph}
\end{figure}

In Fig. \ref{EITcompare} we plot the underlying EIT curves and the noise correlation resonances for side-by-side comparison from the same experiment as that shown in Fig. \ref{g2_two_cells}.  The baseline of the EIT curves have been subtracted so as to easily compare the widths of the curves.  In all cases the noise correlation peak is more narrow than the underlying EIT, as expected.  Also as expected, increasing buffer gas pressure narrows the EIT peak, resulting in a more shallow power broadening rate.  However, the noise correlation resonances show a reverse dependence on pressure.  By increasing buffer gas pressure, the noise correlation resonances broadens and the power broadening rate is increased.  Fig. \ref{summary graph} summarizes these power broadening results.   

This reverse dependence of the $g^2(0)$ noise
correlation linewidth on pressure emerges naturally
from our numerical simulations that model the system
using a three-level atom, as the theory lines in Fig. \ref{summary graph} show.  We note that these theory lines are not curve fits to the data, but instead are representative of our numerical results and are
consistent with observed experimental noise correlation power broadening rates in the buffer gas cells.   
In addition to the typical parameters for ${}^{87}$Rb and the relevant buffer gases, 
observed EIT contrasts are used to determine an optical thickness parameter, used as a simulation input.  

Competition between the optical pumping rate and the ground state
decoherence rate determines the ground state polarization
and thus the robustness of the $g^2(0)$ noise correlation
signal. Further, the excited state relaxation rate and the
reciprocal of the ground state coherence width are proportional to
the buffer gas pressure. Both experiment and theory indicate that
the longer the ground state coherence lifetimes, such as in the 10 Torr cell compared to the 2 Torr cell, 
the more susceptible the atoms are to processes that lead to 
$g^2(0)$ resonance power broadening. The EIT signal is simply not as robust in the 10 Torr cell as in 
the 2 Torr cell, as indicated by the difference in EIT 
contrast in the two cells.  As pointed out in  \cite{YanhongPRL}, laser RIN 
added to the theory as described in \cite{harvard2} and \cite{ariunbold2010intensity} 
can disproportionately affect the weaker EIT system, 
leading to more pronounced $g^2(0)$ noise correlation power broadening. 
The theory lines in Fig. \ref{summary graph} include RIN as well as a single coherence derating parameter as described in Section \ref{sec:gradients}.  This derating parameter conveniently represents the net effect of physical processes, not included in theory models cited above, which depolarize ground state coherence. These depolarizing processes also contribute differentially in buffer gas pressure to the $g^2(0)$ noise correlation power broadening.

While not modeled in our simulations, earlier work on Hanle-configuration EIT in buffer gas cells \cite{Novikova:05} propose that increased collisional broadening of the excited state from increased buffer gas pressure leads to an additional mechanism that depolarizes the ground state coherence.  As the excited state broadens, the F=2 $\rightarrow$ F'=2 hyperfine transition is more likely to be excited along with the adjacent F=2 $\rightarrow$ F'=1 hyperfine transition. Because the ground states are all degenerate at B=0, a double $\Lambda$ system is formed.   The second $\Lambda$ system also supports dark states, but due to a relative sign difference in the dipole matrix elements, those authors suggest that velocity-changing collisions with the buffer gas atoms couples the two $\Lambda$ systems, causing a reduction in ground state coherences and the resulting EIT.  As in their experiments, we observe a decrease in the EIT contrast on the F=2 $\rightarrow$ F'=1 transition with increased buffer gas pressure (see Fig. \ref{EITcompare}).  We also correspondingly observe a decrease in the overall noise amplitude produced by the EIT by a factor of 4.  We do not include the proposed double-$\Lambda$ system cancellation effects in our theory model; however, our simulation results nevertheless show that with increased buffer gas pressure, there is a decrease in EIT contrast and overall noise amplitude, as well as a broadening of the EIT-derived noise correlation linewidth.  To systematically explore the contribution of this double-$\Lambda$ effect, we need to perform a study of noise correlation power broadening as a function of one-photon detuning.   The current work presented in this article focuses exclusively on the F=2 $\rightarrow$ F'=1 transition, or zero one-photon detuning.

\section{Magnetic Field Gradients}
\label{sec:gradients}

The studies described above indicate that the power broadening slope of the EIT-derived noise correlation resonance is greater for a more narrow EIT system.  This indicates that noise correlation resonance broadening 
is enhanced via mechanisms associated with buffer gas pressure.  However, there are other physical processes that also apparently contribute to power broadening of the correlation resonance \cite{mc_so_damop}.  In this section we present another physical process that leads to coherence depolarization, and results in increased power broadening, namely spatial inhomogeneities in the two-photon detuning across the vapor cell.  Via coherence diffusion, these inhomogeneities cause imperfect, bandwidth-limited averaging in the net two-photon detuning, thus leading effectively to ground state coherence depolarization.

To experimentally investigate depolarization due two-photon detuning inhomogeneities, we introduced a controllable $B$-field gradient within the $\mu$-metal shielding using a small current-carrying coil. The coil was placed with its axis parallel to the optical axis. A fluxgate magnetometer was used to map the longitudinal component of the coil's $B$-field gradient within the magnetic shielding for a range of currents through the coil. In the region of the atomic vapor cell, the axial component of the gradient was found to be approximately spatially constant and linear in applied current, as expected.  

\begin{figure}
\includegraphics[width=\linewidth]{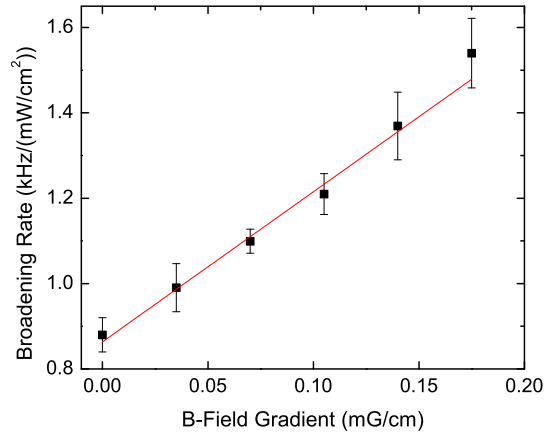}
\caption{(color online) For each applied $B$-field gradient value, the correlation peak FWHM was measured as a function of intensity, and the broadening rate was determined by a linear fit.  For reference, the $0$ mG/cm point was determined by data in Fig.\ \ref{zerogradient_powerbroad}.  Data taken in the 10 Torr vapor cell.} 
\label{gradient_powerbroad}
\end{figure}

For each applied $B$-field gradient value, the correlation peak FWHM was measured as a function of intensity, and the broadening rate was determined by a linear fit, as in Fig.\ \ref{zerogradient_powerbroad}(b).  Fig.\ \ref{gradient_powerbroad} is a plot of the broadening rate of the central correlation peak versus applied $B$-field gradient.  The data suggests that the broadening rate grows linearly with applied $B$-field gradient.  

Simple theory models of EIT noise generally start with analysis of
a single-site three-level system, either treating the time evolution of the 
coherences that contribute to the EIT noise directly or as an ensemble 
average over static responses of such a system,
see \cite{Felinto:13, harvard2, ariunbold2010intensity, LezamaSpectrum}. 
In this manuscript we use the latter approach.   When ensemble averaging over
the fast laser detunings, we fix ground state populations and ground state 
coherences, which have much longer evolution timescales, to their static responses to the average laser 
frequency, as in \cite{harvard2}. 
Single-site theories, however, cannot easily accommodate spatial effects. While random spatial inhomogeneities can be included crudely as contributing to the ground state width, this method is not helpful for modeling the effect of a fixed axial gradient, in which atoms at different locations along the beam's path sample different location-dependent detunings (i.e. longitudinal magnetic fields) during their diffusive transit.

It is illustrative to consider a two-site generalization of the single-site theory. There, in the optically thin limit, the noise contributions from each of the sites along the beam add, but this can be accommodated into an effective single site model in which we average the density matrix components associated with the ground state at the two sites. Thus the populations, as real numbers, average to the same value expected of a single-site theory.  However, because the ground state coherences phase advance at different rates (assumed within an average ground state coherence relaxation time scale), their average is invariably a complex number with a smaller magnitude than what is expected in the single-site theory model. This `derating' of the norm of the ground state coherence reduces the robustness of the $g^2(0)$ signal and in simulation leads directly to a power broadening of the $g^2(0)$ central correlation peak.  For modest deratings of the ground state coherence, simulations indicate a linear relationship between derating and the power broadening slope.  By this multi-site averaging, small B-field gradients cause a derating proportional to the field gradient and inversely proportional to the intrinsic ground state coherence lifetime, consistent with the experimental results in Fig. \ref{summary graph} and Fig. \ref{gradient_powerbroad}. 

To summarize, the experiments suggest that there is no simple way to incorporate the effects of an axial gradient into the single-site theory model for EIT noise correlators through a change in the ground state relaxation parameter, as such a broadening would reduce the power broadening rate. Instead, one may understand the effects seen in experiment as resulting from a reduction in the magnitude of the ground state coherence (relative to the populations) due to effectively averaging  the dephased coherences along the beam. The local ground state coherences along the beam are sampled spatial averages associated with diffusion during the ground state coherence time.

\section{Conclusions}
\label{sec:conclusion}

The intensity noise correlation resonance in EIT-systems illuminated with a low-cost non-narrow linewidth laser 
arises from intrinsic laser phase noise probing ground
state coherence evolution.  Understanding the physical contributions to power broadening of the intensity noise correlation is
technologically relevant. 
Noise spectroscopy techniques could also be used in other
coherent media including EIT in rare-earth doped crystals \cite{rare_earth}, nitrogen-vacancy color
centers in diamond \cite{diamond}, and quantum wells \cite{Wang:12}.

We demonstrate that power broadening of EIT noise 
correlation resonances in a buffer gas cell is a consequence of processes that reduce the 
ground state coherences relative to that expected in the steady 
state for a single atom.  Externally applied noise in the EIT bandwidth, like RIN, 
and imperfect averaging over two-photon detunings caused, for example,
by its gradient are examples of distinct processes that separately contribute to 
noise correlation power broadening.  For fixed RIN, theory and experiments indicate
that the power broadening slope is proportional to the inhomogeneous 
broadening associated with a gradient in the two-photon detuning (see Fig. \ref{gradient_powerbroad})
and is inversely proportional to the intrinsic two-photon width (see Fig. \ref{summary graph}).  Thus our studies reveal a new and counter-intuitive reverse dependence of the
EIT intensity correlation resonance linewidth on buffer gas pressure.  This result is likely to be relevant for EIT noise spectroscopy applications. 

Further studies are underway to systematically disentangle the relative contributions of RIN and coherence derating to power broadening of the noise correlation resonance.  Understanding the relative significance of these two broadening mechanisms for $g^2(0)$ noise correlation power broadening is a prerequisite to using it as a probe of coherence dynamics in the vapor cell.\\


We are grateful to J.\ R.\ Brandenberger, A.\ Leandhardt, I.\ Novikova, D.\ Phillips, 
T.\ Walker,  R.\ Walsworth, and Y.\ Xiao for insightful
discussions and equipment use.  Additionally, SO would like to thank M.\ V.\ Camp for her early and important contributions to this work.

This research was supported by an award from Research Corporation for Science Advancement as well as support under NSF award number 1506499 (SO).  MC acknowledges support under NSF grant number ECCS-
1360725 and from the Science and
Technology Center for Layered Polymeric Systems under
grant number DMR 0423914. 

\bibliography{references}

\end{document}